\documentclass[a4paper,11pt]{article}
\usepackage{pos}

\usepackage{amsmath}
\usepackage{xcolor}

\usepackage{physics}
\newcommand{\1}{\mbox{1}\hspace{-0.25em}\mbox{l}}

\usepackage{graphicx}
\graphicspath{{./figure/}}
\usepackage{tikz}
\usepackage{here}
\usepackage{subcaption}
\usepackage{booktabs}

\title{Numerical simulation of fractional topological charge in $SU(N)$ gauge theory coupled with $\mathbb{Z}_N$ 2-form gauge fields}

\author*[a]{Motokazu Abe}
\author[b]{Okuto Morikawa}

\affiliation[a]{Department of Physics, Kyushu University,
744 Motooka, Nishi-ku, Fukuoka 819-0395, Japan}
\affiliation[b]{Interdisciplinary Theoretical and Mathematical Sciences Program (iTHEMS),
RIKEN, Wako 351-0198, Japan}

\emailAdd{abe.motokazu@phys.kyushu-u.ac.jp}
\emailAdd{okuto.morikawa@riken.jp}

\abstract{
  The pure $SU(N)$ gauge theory with a $\theta$ term has the $\mathbb{Z}_N$ $1$-form global symmetry. When this symmetry is gauged, it is formally established that the topological charge becomes fractional. In this talk, we generate gauge configurations using the HMC method with coupling to the gauged $\mathbb{Z}_N$ $2$-form gauge field. After smoothing these configurations via the gradient flow method, we numerically confirm that the topological charge has a fractional value. We also anticipate that these higher-form fields can solve the topological freezing problem.
}

\FullConference{The 41st International Symposium on Lattice Field Theory (LATTICE2024)\\
 28 July - 3 August 2024\\
Liverpool, UK\\}

\begin{document}
\maketitle

\section{Introduction}
From a modern perspective, the concept of symmetry has been significantly extended and is now referred to as generalized symmetry~\cite{Gaiotto:2014kfa}. For instance, the pure $SU(N)$ gauge theory with $\theta$ term, $\frac{1}{16\pi^2}\theta\int\dd[4]{x} F\wedge F=\theta Q$, where $F$ is the field strength and $Q$ is the topological charge, has the $\mathbb{Z}_N$ $1$-form global symmetry, which is originally known as the center symmetry. Since $Q\in\mathbb{Z}$, the path-integral weight is invariant under $2\pi$ shift of~$\theta$: $e^{-i\theta Q}\to e^{-i\theta Q}$. The $\mathbb{Z}_N$ 1-form global symmetry can be gauged, while $2$-form background gauge fields, $B$, are introduced. Consequently, a mixed 't~Hooft anomaly associated with the higher-form symmetry arises; the key point is that this coupling with $B$ renders $Q$ fractional. It is well known~\cite{Gaiotto:2017yup} that this anomaly is useful to understand the low energy dynamics of gauge fields.

Recently, the fractional topological charge was defined in a completely lattice-regularized framework, and the above mixed 't~Hooft anomaly was proved~\cite{Abe:2022nfq,Abe:2023ncy}. Lattice field theories are fully regularized in an ultra-local and non-perturbative manner, and enable us to perform numerical simulations. In this talk, we numerically compute the fractional topological charge coupled with $B$ fields. Here, we generate those gauge configurations using the Hybrid Monte Carlo (HMC) method with $B$ fields and make each configuration smooth by the gradient flow to compute $Q$.

The coupling with $B$ fields also corresponds to the 't~Hooft twisted boundary condition (b.c.)~\cite{tHooft:1980kjq}. (See also Ref.~\cite{Abe:2023ubg}). Here, we address the well-known issue of topological freezing, a numerical challenge caused by critical slowing down in generating configurations. While the open b.c.\ is an option to address this issue, the 't~Hooft twisted b.c.\ would also resolve this freezing problem. In this talk, to see less severe freezing behaviors, we compare autocorrelation functions for two cases: with $B$-fields ('t~Hooft twisted b.c.) and without them (periodic b.c.).

\section{Lattice setup}\label{sec:numerical_calc}
\subsection{$\mathbb{Z}_N$ 1-form symmetry on the lattice}
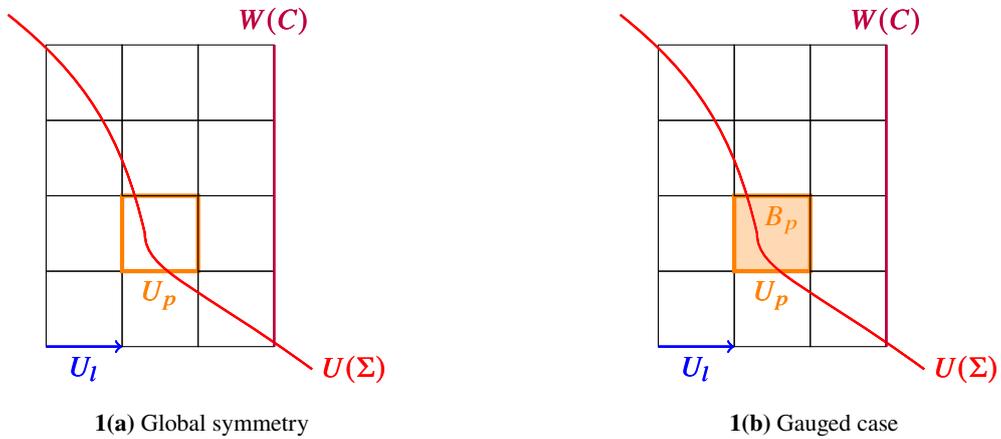
\begin{figure}[htbp]
  \centering
  \begin{minipage}{0.45\textwidth}
      \centering
      \begin{tikzpicture}
          \def\cols{2} 
          \def\rows{3} 
          \def\size{1} 
        
          \foreach \x in {0,...,\cols} {
              \foreach \y in {0,...,\rows} {
                  \ifnum\x=1
                      \ifnum\y=1
                          \draw[orange, ultra thick] (\x*\size,\y*\size) rectangle (\x*\size+\size,\y*\size+\size);
                      \else
                          \draw (\x*\size,\y*\size) rectangle (\x*\size+\size,\y*\size+\size);
                      \fi
                  \else
                      \draw (\x*\size,\y*\size) rectangle (\x*\size+\size,\y*\size+\size);
                  \fi
              }
        
            \draw[red, thick] 
              (-0.5*\size, \rows*\size + 1.4*\size) 
              .. controls (0.5*\size, 3.6*\size) and (1.0*\size, 2.8*\size) 
              .. (1.3*\size, 1.5*\size)
              .. controls (1.3*\size, 1.0*\size) and (2.0*\size, 0.8*\size) 
              .. (\cols*\size + 1.5*\size, -0.3*\size);
        
            \node at (\cols*\size + 1.5*\size, -0.3*\size) [right] {$\textcolor{red}{U(\Sigma)}$};
        
            \node at (1.5*\size, \size) [below] {$\textcolor{orange}{U_p}$};
        
            \draw[blue, thick, ->] (0, 0) -- (\size, 0)
              node[midway, below, black] {$\textcolor{blue}{U_l}$}; 
        
            \draw[purple, thick] (\cols*\size + \size, 0) -- (\cols*\size + \size, \rows*\size + \size);
            \node at (\cols*\size + \size, \rows*\size + \size) [above] {$\textcolor{purple}{W(C)}$};
          }
        \end{tikzpicture}
        \subcaption{Global symmetry}
        \label{fig:ZN_1form_sym_lattice_global}
  \end{minipage}\hspace{3em}
  \begin{minipage}{0.45\textwidth}
      \centering
      \begin{tikzpicture}
          \def\cols{2} 
          \def\rows{3} 
          \def\size{1} 
        
          \foreach \x in {0,...,\cols} {
              \foreach \y in {0,...,\rows} {
                  \ifnum\x=1
                      \ifnum\y=1
                          \fill[orange!30] (\x*\size,\y*\size) rectangle (\x*\size+\size,\y*\size+\size);
                          \draw[orange, ultra thick] (\x*\size,\y*\size) rectangle (\x*\size+\size,\y*\size+\size);
                          \node at (\x*\size+\size, \y*\size+\size) [anchor=north east] {$\textcolor{orange}{B_p}$};
                      \else
                          \draw (\x*\size,\y*\size) rectangle (\x*\size+\size,\y*\size+\size);
                      \fi
                  \else
                      \draw (\x*\size,\y*\size) rectangle (\x*\size+\size,\y*\size+\size);
                  \fi
              }
        
            \draw[red, thick] 
              (-0.5*\size, \rows*\size + 1.4*\size) 
              .. controls (0.5*\size, 3.6*\size) and (1.0*\size, 2.8*\size) 
              .. (1.3*\size, 1.5*\size)
              .. controls (1.3*\size, 1.0*\size) and (2.0*\size, 0.8*\size) 
              .. (\cols*\size + 1.5*\size, -0.3*\size);
        
            \node at (\cols*\size + 1.5*\size, -0.3*\size) [right] {$\textcolor{red}{U(\Sigma)}$};
        
            \node at (1.5*\size, \size) [below] {$\textcolor{orange}{U_p}$};
        
            \draw[blue, thick, ->] (0, 0) -- (\size, 0)
              node[midway, below, black] {$\textcolor{blue}{U_l}$}; 
        
            \draw[purple, thick] (\cols*\size + \size, 0) -- (\cols*\size + \size, \rows*\size + \size);
            \node at (\cols*\size + \size, \rows*\size + \size) [above] {$\textcolor{purple}{W(C)}$};
          }
        \end{tikzpicture}
        \subcaption{Gauged case}
        \label{fig:ZN_1form_sym_lattice_gauge}
  \end{minipage}
  \caption{$\mathbb{Z}_N$ 1-form symmetry transformation on the lattice. The left panel \subref{fig:ZN_1form_sym_lattice_global} illustrates the global symmetry, while the right panel \subref{fig:ZN_1form_sym_lattice_gauge} depicts the gauged symmetry. The red line represents the symmetry operator $U(\Sigma)$, the purple line indicates the Wilson loop $W(C)$, and the orange-shaded region highlights the plaquette coupled with the $B$-fields.}
  \label{fig:ZN_1form_sym_lattice}
\end{figure}

We perform the numerical calculation in the pure $SU(N)$ gauge theory with the following Wilson action:
\begin{align}
  S_W[U_l]
   \equiv \sum_{p \in \mathrm{plaquettes}}\beta\left[ \tr\left( \1-U_p \right) + \mathrm{c.c.} \right],
  \label{eq:dif_wilson_action}
\end{align}
where the considered lattice is $\Lambda = L^4$, $U_l=U_\mu(n)$ denotes link variables living on a bond~$l=(n,\mu)$ with a lattice site~$n\in\Lambda$, $U_p = U_\mu(n) U_\nu(n+\hat{\mu}) U^\dagger_\mu(n+\hat{\nu}) U^\dagger_\nu(n)$ stands for the plaquette term at $p=(n,\mu,\nu)$, and $\beta = 1/g_0^2$. This action is invariant under the $\mathbb{Z}_N$ 1-form global transformation $U_l \to e^{\frac{2\pi i}{N}k}U_l$ ($k=0$, $1$, \dots, $N-1$). Let us define a codimension-$1$ surface $\Sigma$ and symmetry operator on it, $U(\Sigma)$, which acts as the $\mathbb{Z}_N$ 1-form transformation: $U(\Sigma)U_l=e^{\frac{2\pi i}{N}k\mathop{\mathrm{Link}}(\Sigma,l)}U_l$ with an oriented intersection number~$\mathop{\mathrm{Link}}(\Sigma,l)$ between~$\Sigma$ and~$l$. Since $U(\Sigma)$ always passes through a plaquette by an even number of times and $k$ is fixed, $U_p$ is invariant (Fig.~\subref{fig:ZN_1form_sym_lattice_global}). Supposing that $\Sigma$ has a boundary and a loop~$C$ surrounds this codimension-$2$ boundary, the Wilson loop $W(C)$, which is defined by the product of link variables on~$C$, is transformed by $e^{\frac{2\pi i}{N}k}$; that is, the Wilson loop is a charged object associated to the $1$-form symmetry. Moreover, with the knowledge from fiber bundle, gauge transformations between gauge fields in adjacent $4$-dimensional hypercubes, \textit{transition functions $v_{n,\mu}$}, satisfy the cocycle condition $v_{n-\hat{\mu},\nu}(x)v_{n,\mu}(x)v_{n,\nu}(x)^{-1}v_{n-\hat{\nu},\mu}(x)^{-1}=\1$. Here for a general space-time $x\in p=(n,\mu,\nu)$, we can see the explicit form of~$v_{n,\mu}(x)$ in Ref.~\cite{Luscher:1981zq}. Note that this condition is invariant under the $\mathbb{Z}_N$ 1-form symmetry. Next, we gauge this $\mathbb{Z}_N$ 1-form global symmetry. Coupled with 2-form background gauge field $B_p(n)$ to the plaquette, the action becomes 
\begin{align}
  S_W[U_l,B_p]
   \equiv \sum_{p \in \mathrm{plaquettes}}\beta\left[ 
    \tr\left( \1-e^{-\frac{2\pi i}{N}B_p}U_p \right)
     + \mathrm{c.c.}
    \right].
  \label{eq:dif_wilson_action_w_B}
\end{align}
This action is invariant under $U_l\mapsto e^{\frac{2\pi i}{N}\lambda_l}U_l$, $B_p\mapsto B_p+(\dd{\lambda})_p$ (Fig.~\subref{fig:ZN_1form_sym_lattice_gauge}). Then, the cocycle condition is relaxed as following~\cite{Abe:2022nfq,Abe:2023ncy}:
\begin{align}
  \Tilde{v}_{n-\Hat{\nu},\mu}(x) \Tilde{v}_{n,\nu}(x) \Tilde{v}_{n,\mu}(x)^{-1} \Tilde{v}_{n-\Hat{\mu},\nu}(x)^{-1}
   = e^{\frac{2\pi i}{N}B_{\mu\nu}(n-\Hat{\mu}-\Hat{\nu})} \1.
\end{align}

\subsection{Topological charge}
At first, we generate gauge configurations using the HMC method with the action excluding the $B$-fields (Eq.~\eqref{eq:dif_wilson_action}) and calculate the topological charge $Q$ for each configuration. To suppress ultraviolet fluctuations, the link variables are smoothed by the gradient flow method. The flow equation~\cite{Luscher:2010iy} is
\begin{align}
  \partial_t V_t(n,\mu)
  = -g_0^2\left( \partial_{n,\mu}S_W[V_t] \right) V_t(n,\mu),
\quad
  V_t(n,\mu)\big|_{t=0}
  = U_{\mu}(n),
  \label{eq:flow_eq_U}
\end{align}
where $\partial_{n,\mu}$ is the Lie-algebra derivative.
We calculate $Q$ at a specific flow time $t=(0.7L)^2/8$.
By applying the gradient flow method, $Q$ takes an integer value even in numerical lattice calculations. Indeed, many studies have focused on lattice-based instanton calculations employing various actions~\cite{CP-PACS:2001rjn}. Additionally, we adopt an improved method to measure the topological charge $Q$ on the lattice. We consider not only the clover definition~$\overline{Q}^P_L$ but also a rectangular improvement term~$\overline{Q}^R_L$, that is,
\begin{gather}
  Q_{\mathrm{imp}}
    = \sum_{n \in \Lambda} \left\{ c_0\overline{Q}^P_L(n)+c_1\overline{Q}^R_L(n) \right\},
  \label{eq:Q_imp}
  \\
  \overline{Q}^P_{L}(n)
    = \frac{1}{32\pi^2}\epsilon_{\mu\nu\rho\sigma}\Tr(C_{\mu\nu}^P C_{\rho\sigma}^P)
   ,\quad
  \overline{Q}^R_{L}(n) = \frac{2}{32\pi^2}\epsilon_{\mu\nu\rho\sigma}\Tr(C_{\mu\nu}^R C_{\rho\sigma}^R),
  \\
  C_{\mu\nu}^P
    = \frac{1}{4}\Im\left\{
  \begin{tikzpicture}[baseline=(current bounding box.center),scale=0.3]
    \foreach \x in {0,1}{
        \foreach \y in {0,1}{
            \draw (-2+\x*2.1,-2+\y*2.1) rectangle (0++\x*2.1,0+\y*2.1);
          }
      }
    \fill (0.05,0.05) circle (0.2);
  \end{tikzpicture}
  \right\}
  ,\quad
  C_{\mu\nu}^R
    = \frac{1}{8}\Im\left\{
  \begin{tikzpicture}[baseline=(current bounding box.center),scale=0.3]
    \foreach \x in {0,1}{
        \foreach \y in {0,1}{
            \draw (-2+4.1*\x,-2+2.1*\y) rectangle (2+4.1*\x,0+2.1*\y);
          }
      }
    \fill (2.05,0.05) circle (0.2);
  \end{tikzpicture}
  +
  \begin{tikzpicture}[baseline=(current bounding box.center),scale=0.3,rotate=90]
    \foreach \x in {0,1}{
        \foreach \y in {0,1}{
            \draw (-2+4.1*\x,-2+2.1*\y) rectangle (2+4.1*\x,0+2.1*\y);
          }
      }
    \fill (2.05,0.05) circle (0.2);
  \end{tikzpicture}
  \right\},
\end{gather}
where we choose $c_0 = 5/3$ and $c_1 = -1/12$, which is identical to Ref.~\cite{Luscher:1984xn}. 

Next, we generate gauge configurations using an HMC method with the action coupled with $B$-fields (Eq.~\eqref{eq:dif_wilson_action_w_B}). Here, we treat the $B$ field as a dynamical flux rather than a background field, which naively breaks the detailed balance of the HMC method. To restore the detailed balance, we employ a modified HMC algorithm, the ``halfway-updating'' HMC given in Ref.~\cite{Abe:2024fpt}. Since the coupling of the $B$-fields is implemented in the same way as in the action, we substitute $S_W[V_t,B]$ instead of~$S_W[V_t]$ into the flow equation~\eqref{eq:flow_eq_U}; $Q$ acquires fractional values as
\begin{align}
  Q
   =-\frac{1}{N}\int_{T^4}\frac{1}{2}P_2(B_p)\quad\bmod1,
  \label{eq:fractionalcharge}
\end{align}
where the Pontryagin square is defined by
\begin{align}
  P_2(B_p)=B_p\cup B_p+B_p\cup_1\dd{B_p}.
\end{align}
We insert the $B$ field to construct the $\mathbb{Z}_N$ $1$-form symmetry, and it is known that this insertion is equivalent to imposing the 't~Hooft twisted b.c.\ instead of the periodic b.c. This equivalence can be verified under a gauge transformation of the $B$ field. With the 't~Hooft twisted b.c., $Q$ also takes fractional values; we can ensure that the construction in this talk is consistent with Ref.~\cite{vanBaal:1982ag,Abe:2023ubg}.

\subsection{Autocorrelation function and time vs.\ topological freezing}
Physical values generated by the HMC method exhibit autocorrelation, which is a critical metric for obtaining sufficient statistics in numerical calculations. The autocorrelation time of $Q$ increases as the lattice spacing decreases, scaling as $1/a^7$. This phenomenon poses a significant numerical challenge known as topological freezing~\cite{Luscher:2010iy}.

To address this, we first define the autocorrelation function and time. Consider numerical data $\left\{ a_i \right\} (i=1, \dots, N)$ , generated by the Monte Carlo method. Here, these data represent $Q$. The autocorrelation function is defined as
\begin{align}
  \Gamma(\tau) = \frac{1}{N-\tau}\sum^{N-\tau}_{i=1}\left( a_i - \ev{a} \right) \left( a_{i+\tau} - \ev{a} \right),
  \label{eq:autocor_func}
\end{align}
where $\ev{a} = \frac{1}{N}\sum^N_{i=1}a_i$, and $\tau$ is the number of trajectories (molecular dynamics time) in the HMC method. Then, the autocorrelation time $t_{\mathrm{auto}}$ is defined by the normalized autocorrelation function as
\begin{align}
  \rho(\tau) = \Gamma(\tau)/\Gamma(0) \simeq e^{-\tau/t_{\mathrm{auto}}}.
  \label{eq:normalized_Ct}
\end{align}

Furthermore, the integrated autocorrelation time for finite data is 
\begin{align}
  \tau_{\mathrm{int}}(W) = \frac{1}{2} + \sum^W_{i=1}\rho(i),
  \label{eq:autocor_time_finite}
\end{align}
where $W$ is the summation window. If there is no autocorrelation between data, $\tau_{\mathrm{int}} = 1/2$. The stronger the autocorrelation, the larger $\tau_{\mathrm{int}}$ becomes. One solution to mitigate this freezing problem is to impose open b.c.\ instead of periodic b.c.~\cite{Luscher:2011kk}. In this talk, we impose some kind of 't~Hooft twisted b.c.\ through the \textit{random dynamics} of the $B$-fields, and as a result, topological freezing is expected to be resolved.

\subsection{Lattice parameters}
We perform the calculation of the topological charge in the $SU(2)$ and $SU(2)/\mathbb{Z}_2$ gauge theory. Then, we use the following parameters of coupling constant, lattice spacing, and lattice size given in Table~\ref{tab:num_trj} in our actions (Eq.\eqref{eq:dif_wilson_action} and \eqref{eq:dif_wilson_action_w_B}).
\begin{table}[htbp]
  \centering
  \begin{tabular}{llrlc}\toprule
    $\beta$ & $a\sqrt{\sigma}$ & $L$ & $La\sqrt{\sigma}$ & number of configurations
    \\\midrule
    $2.4$ & $0.2673$ & $8$ & $2.138$ & $4000$\\
    $2.5$ & $0.186$ & $12$ & $2.23$ & $4000$\\
    $2.6$ & $0.1326$ & $16$ & $2.122$ & $1000$\\
    \bottomrule
  \end{tabular}
  \caption{Lattice calculations for $SU(2)$ and $SU(2)/\mathbb{Z}_2$ gauge theories and the corresponding number of configurations. For each value of the gauge coupling~$\beta$, the lattice spacing~$a$ is expressed in units of the string tension~$\sigma$, as reported in Ref.~\cite{teper1997physicslatticeglueballsqcd}. Simulations are performed on various lattice sizes, $L^4$, with the corresponding number of configurations listed.}
  \label{tab:num_trj}
\end{table}

\section{Simulation results}
\subsection{Integer topological charge}
\begin{figure}[htbp]
  \centering
  \begin{minipage}{0.45\textwidth}
    \centering
    \includegraphics[keepaspectratio, width=7cm]{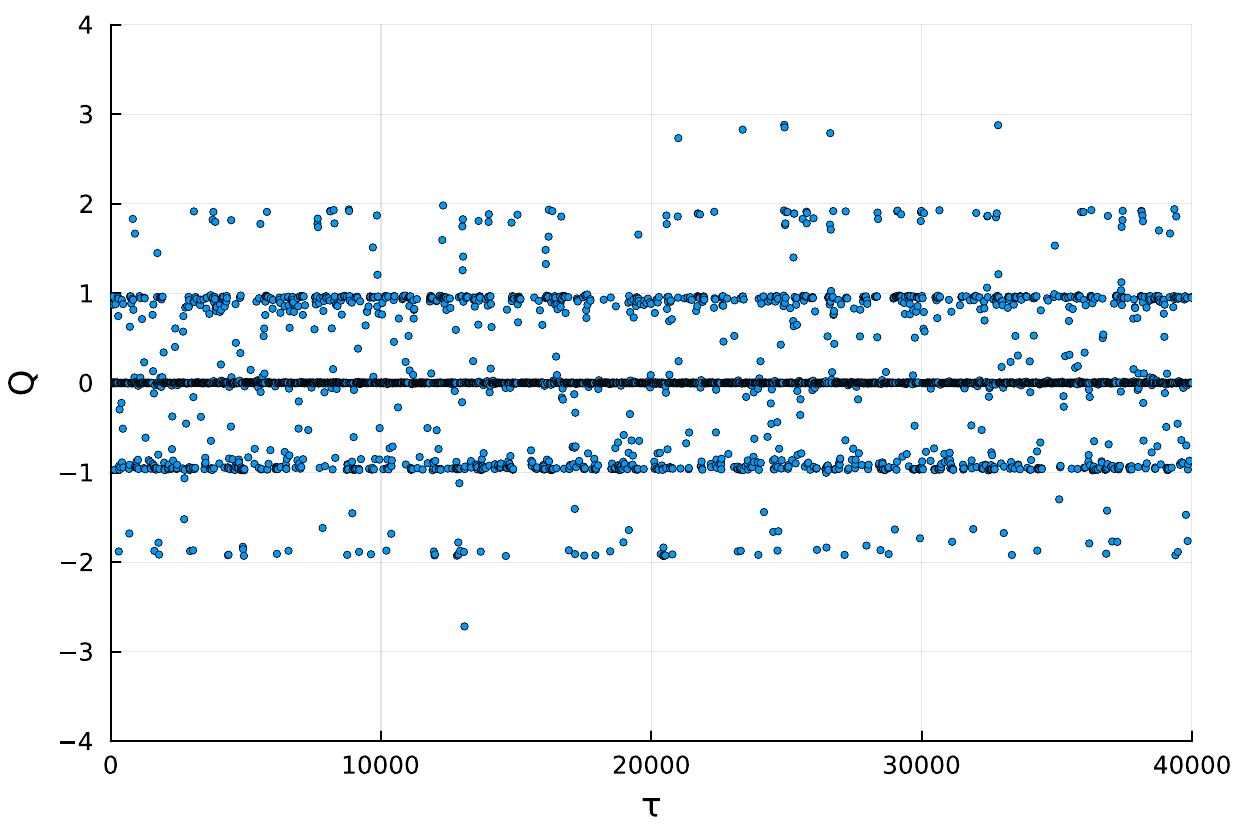}
    \subcaption{$Q$ without the $B$-fields}
    \label{fig:noflux_L8_b2.4_topo}
  \end{minipage}
  \hspace{3em}
  \begin{minipage}{0.45\textwidth}
    \centering
    \includegraphics[keepaspectratio, width=7cm]{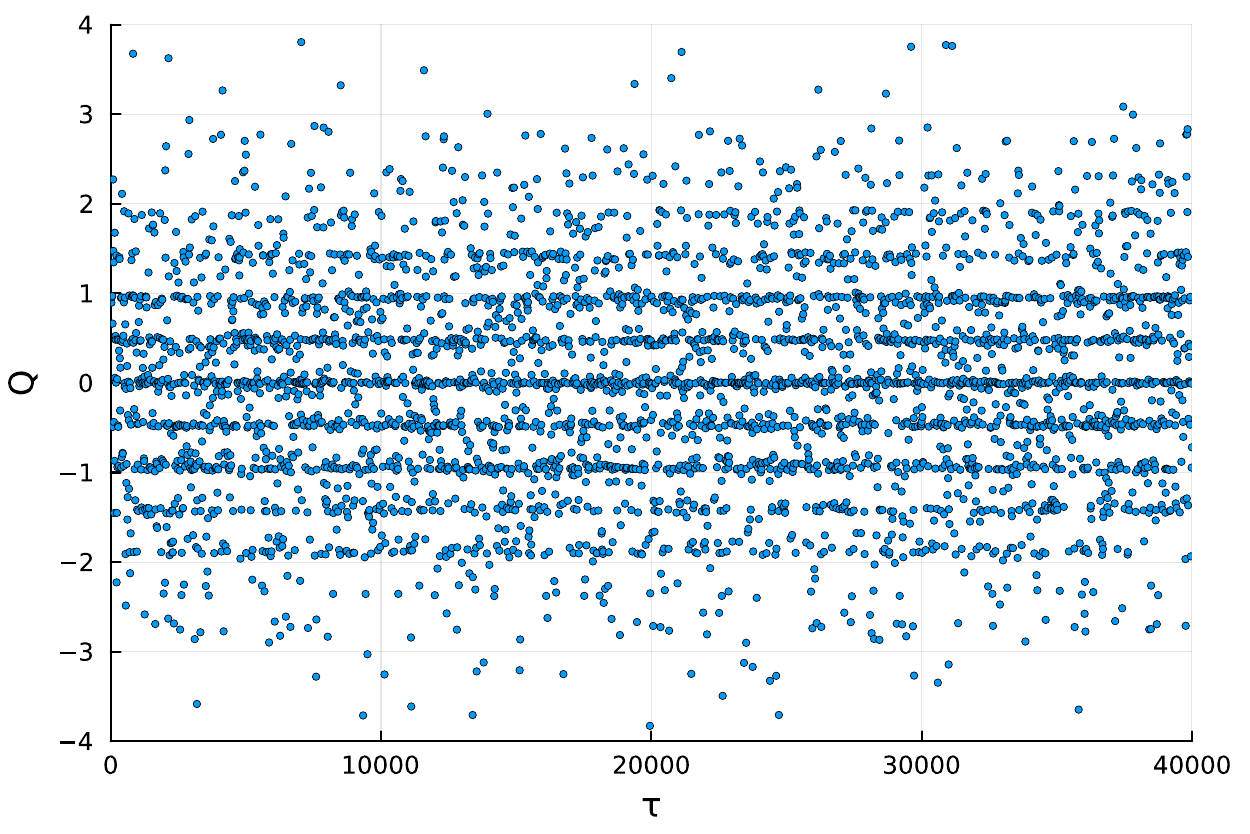}
    \subcaption{$Q$ with $B$ fields}
    \label{fig:beta2.4_L8_with_flux_topo}
  \end{minipage}
  \caption{Topological charge $Q$ vs.\ the trajectory number $\tau$ (molecular dynamics time). Each configuration is separated by $10$ trajectories. The left panel shows that $Q$ is almost integral, while the right panel indicates that $Q$ is fractional. $\beta=2.4$ and $L=8$.}
  \label{fig:topo_vs_config}  
\end{figure}

The result of the calculation of $Q$ using the action without the $B$-fields in Eq.~\eqref{eq:dif_wilson_action}, specifically with $\beta=2.4$ and $L=8$, is shown in Fig.~\subref{fig:noflux_L8_b2.4_topo}. The computation is done for each configuration and depicted with the trajectory number as the horizontal axis. It shows that $Q$ is concentrated around integer values.

\subsection{Fractional topological charge}
\begin{figure}[htbp]
  \centering
  \includegraphics[keepaspectratio, width=7cm]{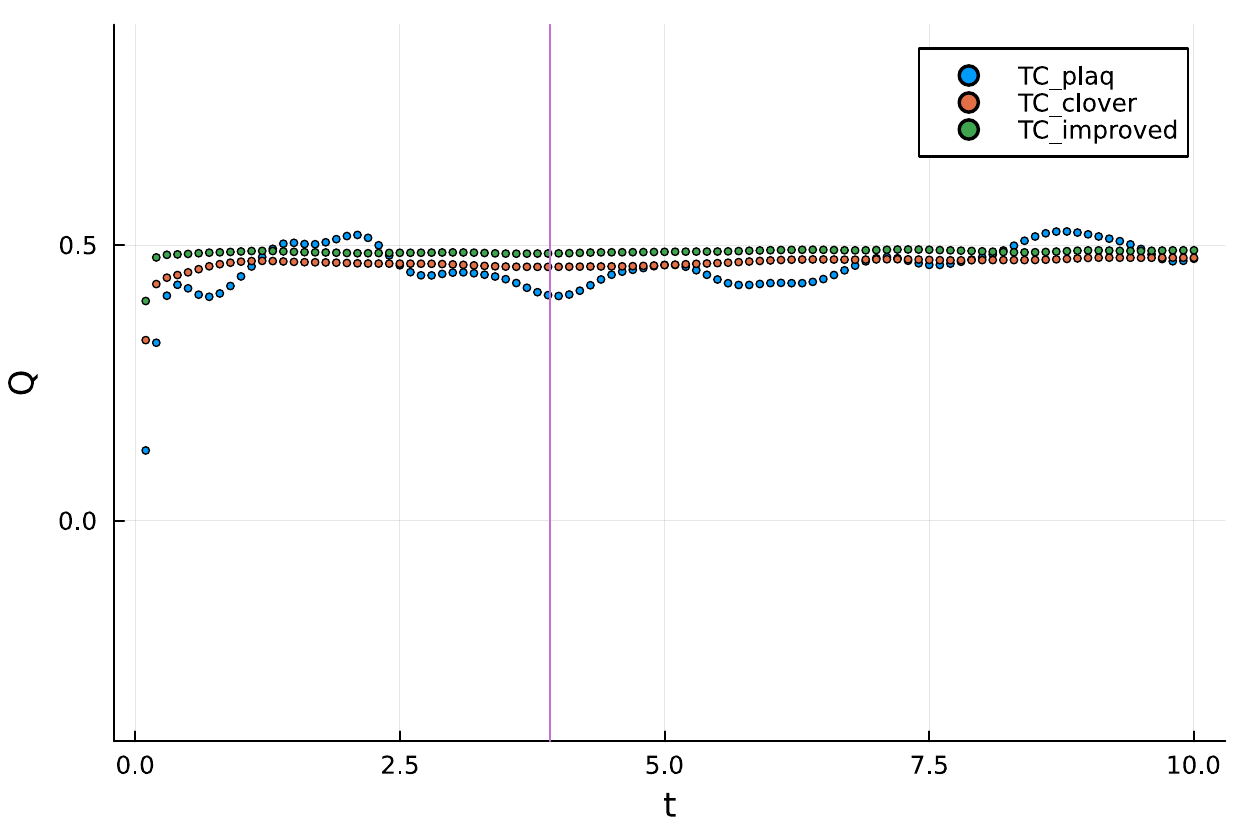}
  \caption{Topological charge $Q$ vs.\ the flow time $t$. The configuration is selected from the region where thermalization has occurred and sufficient time has passed, specifically at $\tau=1000$. From this, it becomes evident that the rectangular improved method we chose exhibits the best convergence. Here, the purple line represents a specific flow time, $t=(0.7L)^2/8$, chosen as the stage before oversmearing in the flow equation~\eqref{eq:flow_eq_U}.}
  \label{fig:TC_glad}
\end{figure}

In Fig.~\subref{fig:beta2.4_L8_with_flux_topo}, we present the results of the calculation of $Q$ using the action coupled with the $B$-fields in Eq.~\eqref{eq:dif_wilson_action_w_B}, where the parameters are set to $\beta=2.4$ and $L=8$. In particular, Fig.~\ref{fig:TC_glad} illustrates how $Q$ evolves with the flow time for a configuration. In this example, $Q$ approaches the fractional value $0.5$ because we set $N=2$ in Eq.~\eqref{eq:fractionalcharge}. We also adapted some definitions of the topological charge: the simple plaquette one, the clover improvement, and our rectangular improved one.
Then, Fig.~\subref{fig:beta2.4_L8_with_flux_topo} shows that $Q$ instead clusters around fractional values, which is multiples of $1/2$ and different from Fig.~\subref{fig:noflux_L8_b2.4_topo}.

\subsection{Normalized autocorrelation function and integrated autocorrelation time}
\begin{figure}[htbp]
  \centering
  \begin{minipage}{0.45\textwidth}
    \centering
    \includegraphics[keepaspectratio, width=7cm]{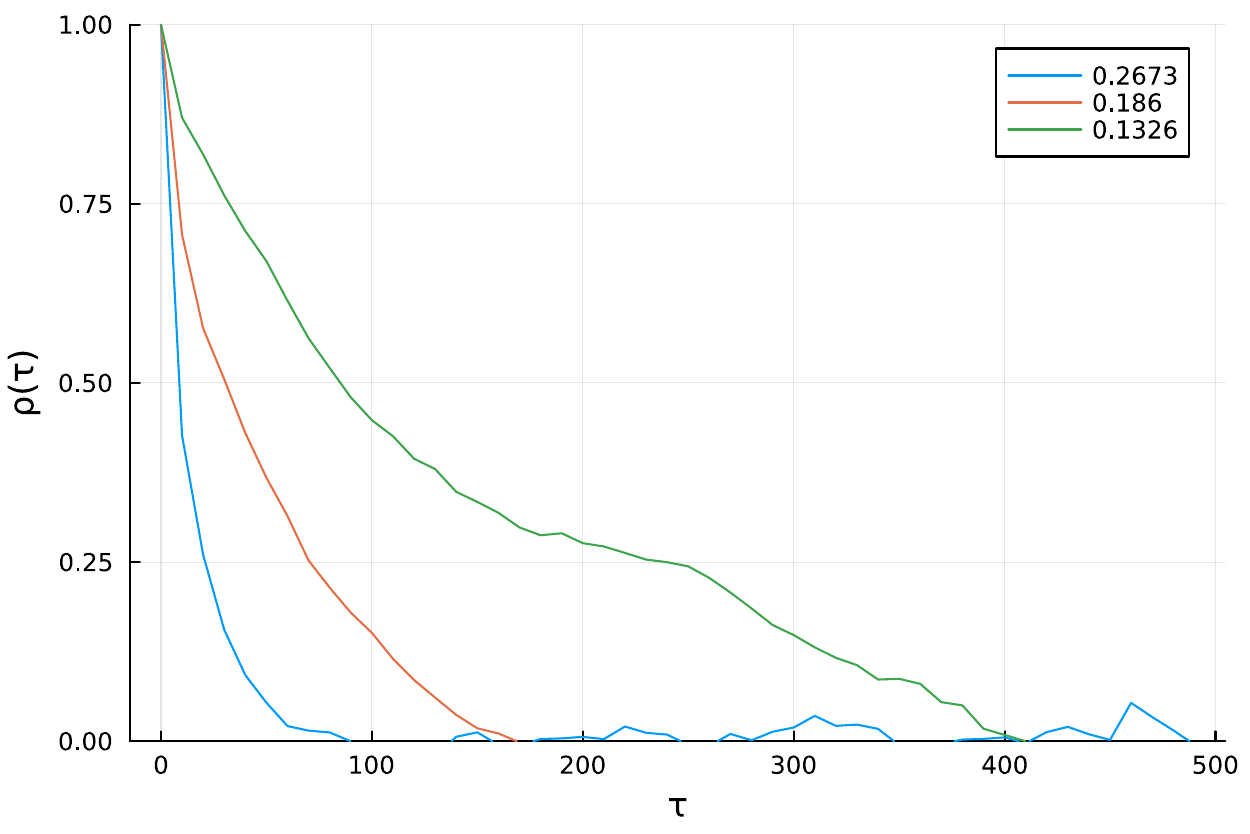}
    \subcaption{$\rho(\tau)$ without the $B$-fields}
    \label{fig:auto_func_wo_flux}
  \end{minipage}
  \hspace{3em}
  \begin{minipage}{0.45\textwidth}
    \centering
    \includegraphics[keepaspectratio, width=7cm]{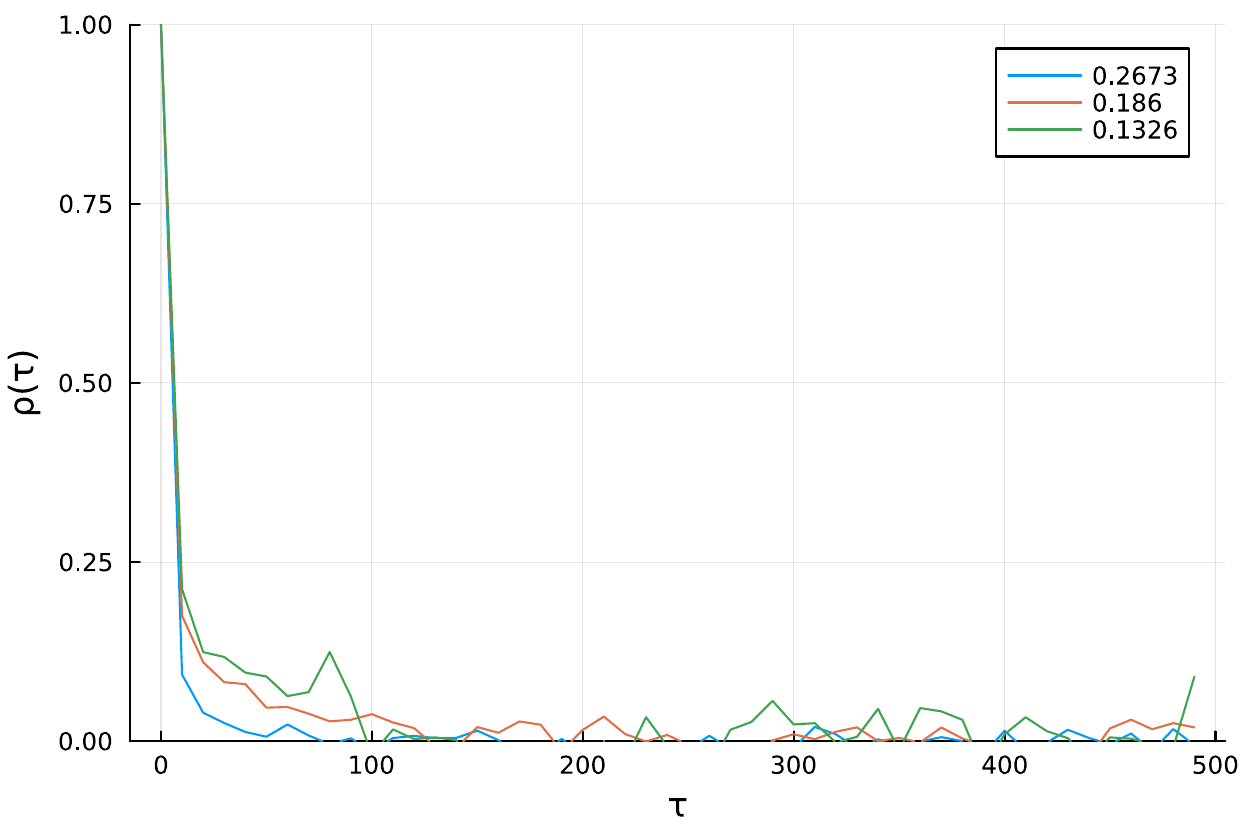}
    \subcaption{$\rho(\tau)$ with the $B$-fields}
    \label{fig:auto_func_w_flux}
  \end{minipage}
  \caption{Normalized autocorrelation function $\rho(\tau)$ of the topological charge $Q$ without (Fig.~\subref{fig:auto_func_wo_flux}) and with (Fig.~\subref{fig:auto_func_w_flux}) the $B$-fields for each $a\sqrt{\sigma}$. The label shows the value of $a\sqrt{\sigma}$ in Table~\ref{tab:num_trj}. From the top of the label, we use the lattice size $L=8,12,16$.}
  \label{fig:auto_func}  
\end{figure}
From Fig.~\ref{fig:auto_func}, we see the normalized autocorrelation function $\rho(\tau)$. The autocorrelation time increases as the lattice spacing $a$ becomes larger. However, when the $B$-fields are introduced, the autocorrelation time becomes remarkably smaller. This suggests that topological freezing is mitigated by imposing the 't~Hooft twisted b.c.

\section{Conclusion and future works}
In this talk, we numerically calculate the fractional topological charge by introducing the $B$-fields, which were formally constructed on the lattice~\cite{Abe:2022nfq,Abe:2023ncy}. We also compute the autocorrelation time in the presence of the $B$-fields and demonstrate that the issue of topological freezing is mitigated by imposing the 't~Hooft twisted boundary condition.

The other research in Ref.~\cite{Luscher:2011kk}, which addresses topological freezing by imposing open boundary conditions, shows that the integrated autocorrelation time scales as $1/a^2$. In our setup, we aim to verify the behavior of the integrated autocorrelation time as a function of the lattice spacing $a$ by increasing the number of data points at different values of $a$.

Furthermore, by introducing the $B$-fields, we plan to numerically calculate the 't~Hooft partition function as discussed in Ref.~\cite{Nguyen:2023fun}. Since this partition function serves as an order parameter for gauge field confinement, we aim to confirm the phase diagram of the $SU(2)/\mathbb{Z}_2$ gauge theory through numerical calculations. In the $SU(2)$ gauge theory, quite recently, the computation of the 't~Hooft partition function using the Monte Carlo method is addressed in Ref.~\cite{Morikawa:2025ldq}.

\subsection*{Acknowledgements}
The numerical calculation of this work is partially performed by Hiroshi Suzuki. Our numerical codes can be found in \url{https://github.com/o-morikawa/Gaugefields.jl}, which is
an extension of Gaugefields.jl in the JuliaQCD package~\cite{Nagai:2024yaf}. This work was partially supported by Japan Society for the Promotion of Science (JSPS) Grant-in-Aid for Scientific Research Grant Numbers JP21J30003, JP22KJ2096 (O.M.). O.M.\ acknowledges the RIKEN Special Postdoctoral Researcher Program. The work of M.A.\ was supported by Kyushu University Innovator Fellowship Program in Quantum Science Area.

\bibliographystyle{JHEP}
\bibliography{ref}

\providecommand{\href}[2]{#2}\begingroup\raggedright\begin{thebibliography}{10}

\bibitem{Gaiotto:2014kfa}
D.~Gaiotto, A.~Kapustin, N.~Seiberg and B.~Willett, \emph{{Generalized Global Symmetries}}, \href{https://doi.org/10.1007/JHEP02(2015)172}{\emph{JHEP} {\bfseries 02} (2015) 172}.

\bibitem{Gaiotto:2017yup}
D.~Gaiotto, A.~Kapustin, Z.~Komargodski and N.~Seiberg, \emph{{Theta, Time Reversal, and Temperature}}, \href{https://doi.org/10.1007/JHEP05(2017)091}{\emph{JHEP} {\bfseries 05} (2017) 091}.

\bibitem{Abe:2022nfq}
M.~Abe, O.~Morikawa and H.~Suzuki, \emph{{Fractional topological charge in lattice Abelian gauge theory}}, \href{https://doi.org/10.1093/ptep/ptad009}{\emph{PTEP} {\bfseries 2023} (2023) 023B03} [\href{https://arxiv.org/abs/2210.12967}{{\ttfamily 2210.12967}}].

\bibitem{Abe:2023ncy}
M.~Abe, O.~Morikawa, S.~Onoda, H.~Suzuki and Y.~Tanizaki, \emph{{Topology of SU(N) lattice gauge theories coupled with \ensuremath{\mathbb{Z}}$_{N}$ 2-form gauge fields}}, \href{https://doi.org/10.1007/JHEP08(2023)118}{\emph{JHEP} {\bfseries 08} (2023) 118} [\href{https://arxiv.org/abs/2303.10977}{{\ttfamily 2303.10977}}].

\bibitem{tHooft:1980kjq}
G.~'t~Hooft, \emph{{Confinement and Topology in Nonabelian Gauge Theories}}, {\emph{Acta Phys. Austriaca Suppl.} {\bfseries 22} (1980) 531}.

\bibitem{Abe:2023ubg}
M.~Abe, O.~Morikawa and S.~Onoda, \emph{{Note on lattice description of generalized symmetries in $SU(N)/\mathbb{Z}_N$ gauge theories}}, \href{https://doi.org/10.1103/PhysRevD.108.014506}{\emph{Phys. Rev. D} {\bfseries 108} (2023) 014506} [\href{https://arxiv.org/abs/2304.11813}{{\ttfamily 2304.11813}}].

\bibitem{Luscher:1981zq}
M.~L{\"u}scher, \emph{{Topology of Lattice Gauge Fields}}, \href{https://doi.org/10.1007/BF02029132}{\emph{Commun. Math. Phys.} {\bfseries 85} (1982) 39}.

\bibitem{Luscher:2010iy}
M.~L\"uscher, \emph{{Properties and uses of the Wilson flow in lattice QCD}}, \href{https://doi.org/10.1007/JHEP08(2010)071}{\emph{JHEP} {\bfseries 08} (2010) 071} [\href{https://arxiv.org/abs/1006.4518}{{\ttfamily 1006.4518}}].

\bibitem{CP-PACS:2001rjn}
{\scshape CP-PACS} collaboration, \emph{{Topological susceptibility in lattice QCD with two flavors of dynamical quarks}}, \href{https://doi.org/10.1103/PhysRevD.64.114501}{\emph{Phys. Rev. D} {\bfseries 64} (2001) 114501} [\href{https://arxiv.org/abs/hep-lat/0106010}{{\ttfamily hep-lat/0106010}}].

\bibitem{Luscher:1984xn}
M.~L\"uscher and P.~Weisz, \emph{{On-shell improved lattice gauge theories}}, \href{https://doi.org/10.1007/BF01205792}{\emph{Commun. Math. Phys.} {\bfseries 98} (1985) 433}.

\bibitem{Abe:2024fpt}
M.~Abe, O.~Morikawa and H.~Suzuki, \emph{{Monte Carlo simulation of the $SU(2)/\mathbb{Z}_2$ Yang--Mills theory}},  \href{https://arxiv.org/abs/2501.00286}{{\ttfamily 2501.00286}}.

\bibitem{vanBaal:1982ag}
P.~van Baal, \emph{{Some Results for SU(N) Gauge Fields on the Hypertorus}}, \href{https://doi.org/10.1007/BF01403503}{\emph{Commun. Math. Phys.} {\bfseries 85} (1982) 529}.

\bibitem{Luscher:2011kk}
M.~L\"uscher and S.~Schaefer, \emph{{Lattice QCD without topology barriers}}, \href{https://doi.org/10.1007/JHEP07(2011)036}{\emph{JHEP} {\bfseries 07} (2011) 036} [\href{https://arxiv.org/abs/1105.4749}{{\ttfamily 1105.4749}}].

\bibitem{teper1997physicslatticeglueballsqcd}
M.~Teper, \emph{Physics from the lattice: glueballs in qcd; topology; su(n) for all n},  1997.

\bibitem{Nguyen:2023fun}
M.~Nguyen, Y.~Tanizaki and M.~\"Unsal, \emph{{Study of gapped phases of 4d gauge theories using temporal gauging of the~$\mathbb{Z}_N$ 1-form symmetry}}, \href{https://doi.org/10.1007/JHEP08(2023)013}{\emph{JHEP} {\bfseries 08} (2023) 013} [\href{https://arxiv.org/abs/2306.02485}{{\ttfamily 2306.02485}}].

\bibitem{Morikawa:2025ldq}
O.~Morikawa and H.~Suzuki, \emph{{Direct Monte Carlo computation of the 't~Hooft partition function}},  \href{https://arxiv.org/abs/2501.07042}{{\ttfamily 2501.07042}}.

\bibitem{Nagai:2024yaf}
Y.~Nagai and A.~Tomiya, \emph{{JuliaQCD: Portable lattice QCD package in Julia language}},  \href{https://arxiv.org/abs/2409.03030}{{\ttfamily 2409.03030}}.

\end{thebibliography}\endgroup

\end{document}